\documentclass[preprint,3p,twocolumn,authoryear]{elsarticle}
\usepackage{graphicx,txfonts}
\usepackage{caption,listings}
\usepackage[polish,english]{babel}

\newcommand\code[1]{\texttt{#1}}
\newcommand\mechanic{\texttt{Mechanic}}
\newcommand\mcap[1]{{\bfseries #1}}

\def\v#1{{{\bf #1}}}

\def\mechanic{{\tt Mechanic}}

\def\norm#1{|\,#1\,|}

\journal{New Astronomy}

\begin{document}

\lstdefinelanguage{mechanic}
  {morekeywords={
    Init,PoolPrepare,PoolProcess,TaskPrepare,TaskProcess,
    Setup,Storage,MReadOption,MWriteOption,MReadData,MWriteData,
    MReadAttr,MWriteAttr,MAllocate2,MAllocate3,MAllocate4,MGetDims,
  },
  sensitive=true,
  morecomment=[l]{//},
  morecomment=[s]{/*}{*/},
  morestring=[b]",
  }

\lstset{
  language=mechanic,
  basicstyle=\small\ttfamily,
  frame=t,
  framesep=1.0\baselineskip,
  framerule=0.2pt,
  tabsize=2,
  columns=flexible,
  keepspaces=true,
  title=\lstname,
  showspaces=false,
  showstringspaces=false,
  escapeinside={@}{@},
  emph={Setup,MReadOption,MWriteOption},
  emphstyle=\bfseries,
  float=*,
  floatplacement=t
}

\begin{frontmatter}

\title{Mechanic: the MPI/HDF code framework for dynamical astronomy}
\author{Mariusz S\l{}onina\fnref{ms}, Krzysztof Go\'zdziewski\fnref{kg}, and
Cezary Migaszewski\fnref{cm}}

\fntext[ms]{E-mail: m.slonina@astri.umk.pl}
\fntext[kg]{E-mail: k.gozdziewski@astri.umk.pl}
\fntext[cm]{E-mail: c.migaszewski@astri.umk.pl}
\address{Toru\'n Centre for Astronomy, Nicolaus Copernicus University, Gagarin Str. 11,
87-100 Torun, Poland}

\begin{abstract}
We introduce the \mechanic{}, a new open-source code framework. It is designed to reduce the
development effort of scientific applications by providing unified API (Application
Programming Interface) for configuration,
data storage and task management. The communication layer is based on the well-established Message Passing
Interface (MPI) standard, which is
widely used on variety of parallel computers and CPU-clusters. 
The data storage is performed within the Hierarchical Data Format (HDF5). The design of
the code follows {\em core--module} approach which allows to reduce the user's codebase
and makes it portable for single- and multi-CPU environments. The framework may be used in
a local user's environment, without administrative access to the cluster, under the PBS or
Slurm job schedulers.
It may become a helper tool for a wide range of
astronomical applications, particularly focused on processing large data sets, such as dynamical studies
of long-term orbital evolution of planetary systems with Monte Carlo methods, dynamical
maps or evolutionary algorithms. It has been already applied in numerical
experiments conducted for Kepler-11 \citep{kepler-11} and $\nu$Octantis planetary systems \citep{nuoct}. 
In this paper we describe the basics of the framework, including code listings for the
implementation of a sample user's module. 
The code is illustrated on a model Hamiltonian introduced by \citep{Froeschle2000}
presenting the Arnold diffusion. The Arnold Web is shown with the help
of the MEGNO (Mean Exponential Growth of Nearby Orbits) fast indicator \citep{Gozdziewski2008} 
applied onto symplectic SABA$_n$ integrators family \citep{Laskar2001}.
\end{abstract}


\end{frontmatter}

\section{Introduction} 
In the field of dynamical astronomy several numerical techniques have been proposed 
to determine the nature of the phase space of planetary systems.
The Monte Carlo methods \citep[e.g.,][]{Holman1999},
evolutionary algorithms \citep[e.g.,][]{gamp2008,gamp2009} or dynamical maps
\citep[e.g.,][]{Froeschle2000,Guzzo2005,kepler-11,nuoct} 
have become standard research tools for determining
possible or permitted configurations, mass ranges or other physical data. 
These experiments usually require intensive tests of sets of initial conditions,
that represent different orbital configurations.
They involve
direct numerical integrations of equations of motion to study long-term orbital evolution.
To characterize the dynamical stability of orbital models, so called {\em fast 
chaos indicators} are often used \citep[e.g.,][]{Gozdziewski2008}. These numerical 
tools make it possible to resolve efficiently whether a given solution is 
stable (quasi-periodic, regular) or unstable (chaotic) by following 
relatively short parts of the orbits. The fast indicators, like the Fast 
Lyapunov Indicator \citep [FLI,][]{Froeschle2000}, the Frequency Map 
Analysis \citep [FMA,][]{Laskar1993,Nesvorny1996}, the Mean Exponential 
Growth factor of Nearby Orbits \citep 
[MEGNO,][]{cs2000,Cincotta2003,Mestre2011}, the Spectral Number \citep 
[SN,][]{Michtchenko2001}, are well known in the theory of dynamical systems 
\citep{Barrio2009}. In the past decade, they were intensively adapted to 
the planetary dynamics \citep[e.g.,][]{Froeschle1997,Robutel2001,Gozdziewski2008}.

Depending on the dynamical model of a planetary system, its
numerical setup and the chaos indicator used to represent the dynamical state, the
simulation of a set of initial conditions may require large CPU resources. 
However, since
each test may be understood as a separated numerical {\em task}, 
parallelization techniques may be used, with tasks distributed among available CPUs and
evaluated in {\em parallel}.
The basic approach relies on the {\em task farm} model, in which independent
tasks are processed on {\em worker} nodes with the result collected by the {\em master}.
From the technical point of view, this algorithm requires implementation of the
CPU-communication layer, and should allow 
input preparation and result assembly for the post-processing. These issues are
addressed in general-purpose distributed task management systems, 
like \code{HTCondor} \citep{condor}, or \code{Workqueue} \citep{workqueue}. 
Within such frameworks, the user-supplied, standalone executable code performing
computations ({\em application}) is distributed over a computing pool. 
The input and output for each software instance is achieved via batch scripts 
(e.g. the \code{Makeflow} extension for the \code{Workqueue} package),
making this approach application- and problem-dependent.
This might be insufficient for large and long-term numerical tests, such as studying the
dynamics of planetary systems. In particular, our recent work on Kepler-11
\citep{kepler-11} and $\nu$Octantis \citep{nuoct} systems required developing a new
code framework, the \mechanic{}, dedicated to conducting massive parallel simulations. It has been
turned out into general-purpose master--worker framework, built on the foundation of the
Message Passing Interface \citep{Pacheco}. The \mechanic{} separates the
numerical part of the user's code ({\em a module}) from its configuration, communication
and storage layers ({\em a core}). This partition is achieved through 
the provided Application Programming Interface (API).
On the contrary to \code{HTCondor} and \code{Workqueue}
packages, the task preparation and result data storage is handled by the core of the
framework. The final result is assembled into one datafile, which reduces the cost of
post-processing large simulations. The storage layer is built on top of the universal HDF5
data format \citep{hdf}. No MPI nor HDF5 programming knowledge is required to use the
framework, which makes it possible to parallelize ``scalar'' codes relatively easily. The
\mechanic{} may be used both system-wide as well as in a local user's environment under
the control of job schedulers, such as PBS or Slurm.

This paper is structured as follows. We give a short overview of the \mechanic{} 
in Section \ref{sec:overview}. To explain programming concepts behind the framework we 
illustrate it with the help of the Hamiltonian model introduced by
\cite{Froeschle2000}. It reveals the so called {\em Arnold web}, which represents a set of
resonances of a quasi-integrable dynamical systems. It has been intensively
studied in recent years
\citep{Cincotta2002,Lega2003,Guzzo2004,Froeschle2005,Froeschle2006}, and applied to study
long-term evolution of the outer Solar System \citep{Guzzo2005,Guzzo2006}.
In Section \ref{sec:arnold} we give short theoretical background on this topic. 
The very fine details of the phase space obtained with the dedicated module for the
\mechanic{} are presented in the Section \ref{sec:results}. 
The technical implementation of the module is given 
in the Appendix.

\section{Overview of the framework}\label{sec:overview} 
The \mechanic{} provides a skeleton code
for common technical operations, including run-time configuration, memory and
file management, as well as CPU communication. It has been developed to mimic the
user's application flow in a problem-independent way (Listing~\ref{lst:design}). This is achieved via
provided API, which allows to reduce the user's code to a {\em module} form,
containing only its numerical part along with setup and storage specifications
required to run it (Listing~\ref{lst:module}). The {\em core} of the framework loads the module dynamically
during the runtime, performs the setup and storage stages according to these specs
and executes the numerical part.

The benefit of this {\em core--module} approach comes both in data and task
management. For instance, let us recall the concept of dynamical maps. The phase space of
the dynamical system is mapped onto two-dimensional plane. Each point on that plane
represents the dynamical state of the specific initial condition.
From the technical point of view, computing the dynamical map requires
execution of several numerical tasks that differs with the input, and assembling the
result in an accessible way for post-processing. Assuming that each task (initial
condition) is computed by a single instance of the application, the simulation
requires preparing the input and collecting the result with the help of batch
scripts. Although the \code{HTCondor} and \code{Workqueue} frameworks provide powerful
task management tools, input and output data management is left to the user. 
The \mechanic{} framework works more like \code{Makeflow} (a \code{Workqueue} make
engine), however, the user's code
connected to the core is treated as a whole application with single output datafile
and the input that may be prepared programatically according to the information
associated with the current task. 

\begin{lstlisting}[caption={\mcap{The internal design of the framework (pseudocode).}
The user's code is
connected with the framework through the provided API (hooks are marked with bold font
face). API hooks used in the module are executed in a specific order to mimic the user's
application flow. The bootstrap stage
involves \code{Init()} and \code{Setup()} hooks for code initialization. After that, the framework
enters the task pool loop. For each task pool \code{p}, the information provided with the
\code{Storage()}
hook is used to allocate the required memory and the file storage. The \code{PoolPrepare()} hook
allows to prepare the global data and configuration of the pool \code{p}. This hook has access to the data computed
in the previous \code{pools}, if any. Each task pool \code{p} involves evaluation of the
task loop. The set of numerical tasks is allocated on the core part, so that they are
available for use in the API. For each task, the
\code{TaskPrepare()} and \code{TaskProcess()} hooks are executed. The computed data is
saved through the task \code{t} object that is passed to the hooks. In the
case of the master--worker approach, the \code{for} loop showed in the listing is
parallelized. The task \code{t} is initialized on the master node and sent to the worker.
\code{TaskPrepare()} and \code{TaskProcess()} hooks are executed on the worker node, and the data of the task \code{t} is passed back to
the master node and saved. After the task loop is completed, the \code{PoolProcess()} hook is used
to determine whether to continue the task pool loop. To do so, the user may use the data stored in
the previous \code{pools}, if any. During the framework execution, memory and file management,
CPU-communication and data storage are performed on the core part without the need of user
interaction.},label={lst:design}]
Init()
Setup()
  
allocate task pools
 
// task pool loop:
do {
  allocate new task pool p
  Storage(p)
  
  allocate tasks in the task pool p
  PoolPrepare(pools, p)

  // task loop:
  for all t in tasks {
    get the task t
    TaskPrepare(p, t)
    TaskProcess(p, t)
    save the task t
  }
      
  new = PoolProcess(pools, p)
  save task pool p in pools
} while (new == POOL_CREATE_NEW)
\end{lstlisting}

\begin{lstlisting}[caption={\mcap{The sample module structure.} 
Each API hook is provided with the information that is
neccessary on the corresponding execution stage, such as pool and task configuration,
through \code{pool *p} and \code{task *t} pointers. The default return code
\code{SUCCESS} indicates the successful evaluation of the hook. Otherwise, the hook
must return error code, as specified in the API documentation.
This helps the framework to safely abort the simulation. The listing shows only the most
essential API hooks. See the package documentation for the detailed list of
hooks.},label={lst:module}]
#include "mechanic.h"

int Init(init *i) {
  return SUCCESS;
}

int Setup(setup *s) {
  return SUCCESS;
}

int Storage(pool *p) {
  return SUCCESS;
}

int PoolPrepare(pool **pools, pool *p) {
  return SUCCESS;
}

int TaskPrepare(pool *p, task *t) {
  return SUCCESS;
}

int TaskProcess(pool *p, task *t) {
  return TASK_FINALIZE;
}

int PoolProcess(pool **pools, pool *p) {
  return POOL_FINALIZE;
}

\end{lstlisting}

The base master--worker algorithm with single input and output may be easily implemented
with the minimum knowledge on the MPI programming. However, the purpose of the \mechanic{}
is to reduce this development effort. With the help of the API, the user's code 
is separated from the task management layer. It allows to
use different communication patterns between nodes in a computing pool (cluster)
without modification of the module. In addition to the master--worker pattern, the
master-only mode without MPI communication is provided, which behaves similar to a
single-CPU application. Moreover, the API allows to implement different
communication patterns, if required by the user's code. 

The task assignment is performed within multidimensional grid and
is governed through the API. Although this suits best the concept of dynamical maps, the
API has been designed to support different assignment patterns, such as Monte Carlo
methods. The key design concept of the \mechanic{} is a {\em task pool}.
It represents a set of numerical tasks to perform for a particular setup (i.e. single
dynamical map). The framework allows to create task pools dynamically depending on the
results, with different configuration, storage and number of tasks. This helps to
implement evolutionary algorithms and processing pipelines despite of number of CPUs
involved in the simulation.

The result data, among with the run-time configuration, is assembled into one master datafile.
Each task may hold unlimited number of multidimensional arrays of all native datatypes.
Depending on the application requirements, the results obtained from all tasks may be
combined in different modes. 
This includes {\em texture}, represented by a single
dataset that follows the grid pattern (suitable for image-like results, such as dynamical maps), {\em group}
(datasets are combined into separate groups per task), {\em list} (a
spreadsheet-like dataset) and {\em pm3d} (dataset prepared for \code{pm3d} mode of
Gnuplot). The usage of the single master file helps to reduce the post-processing effort. 
Numerous HDF-oriented applications are
available (such as \code{h5py}), making it possible to use computation results
independently of the host software.

For long-term simulations the checkpoint feature is important. Indeed, during the
computations, the \mechanic{} provides the incremental snapshot file with the current
state of the simulation. The file is self-contained, and includes run-time 
configuration, so that no other information is required to restart the job.
By default, only the
evaluated tasks are kept, however the API allows to keep {\em intermediate snapshots} of
each task, containing i.e. temporary simulation data or time snapshots. The working
implementation of this feature is provided with the sample module in the Appendix. 

The framework is developed in a reliable compromise between flexibility and the code
performance. The total amount of RAM and hard drive
storage that is requested by the framework during the simulation depends on the
applications requirements. 
The minimum memory footprint of the core is ensured, so that the full host
resources are available for the user's module. The main concern of the scalability of the
framework is the performance of the MPI communication pattern used during the task
distribution. As a proof-of-concept, the framework uses the MPI-blocking master--worker
communication type. It was primarily designed to suit large and long-term simulations. 
Therefore, it may become a bottleneck for large and fast simulations. The development of
non-blocking task farm, as well as research on different task distribution patterns are
the subject of the follow-up work.

The \mechanic{} is developed in C and supports any C-interoperable programming languages,
including Fortran2003+. It runs on any UNIX-like operating
system (Linux and MAC OS X are actively maintained) and works uniformly in single- and
multi-CPU environments. The framework may be used in a local user's environment, without
administrative access to the cluster.
The package ships with the simple template modules for creating dynamical maps, using
evolutionary algorithms, as well as input data preprocessing and connecting Fortran codes.
They are available at the project page\footnote{http://github.com/mslonina/mechanic}.

\begin{figure*}
\centerline{
\vbox{
 \hbox{
   \includegraphics[     width=0.499\textwidth]{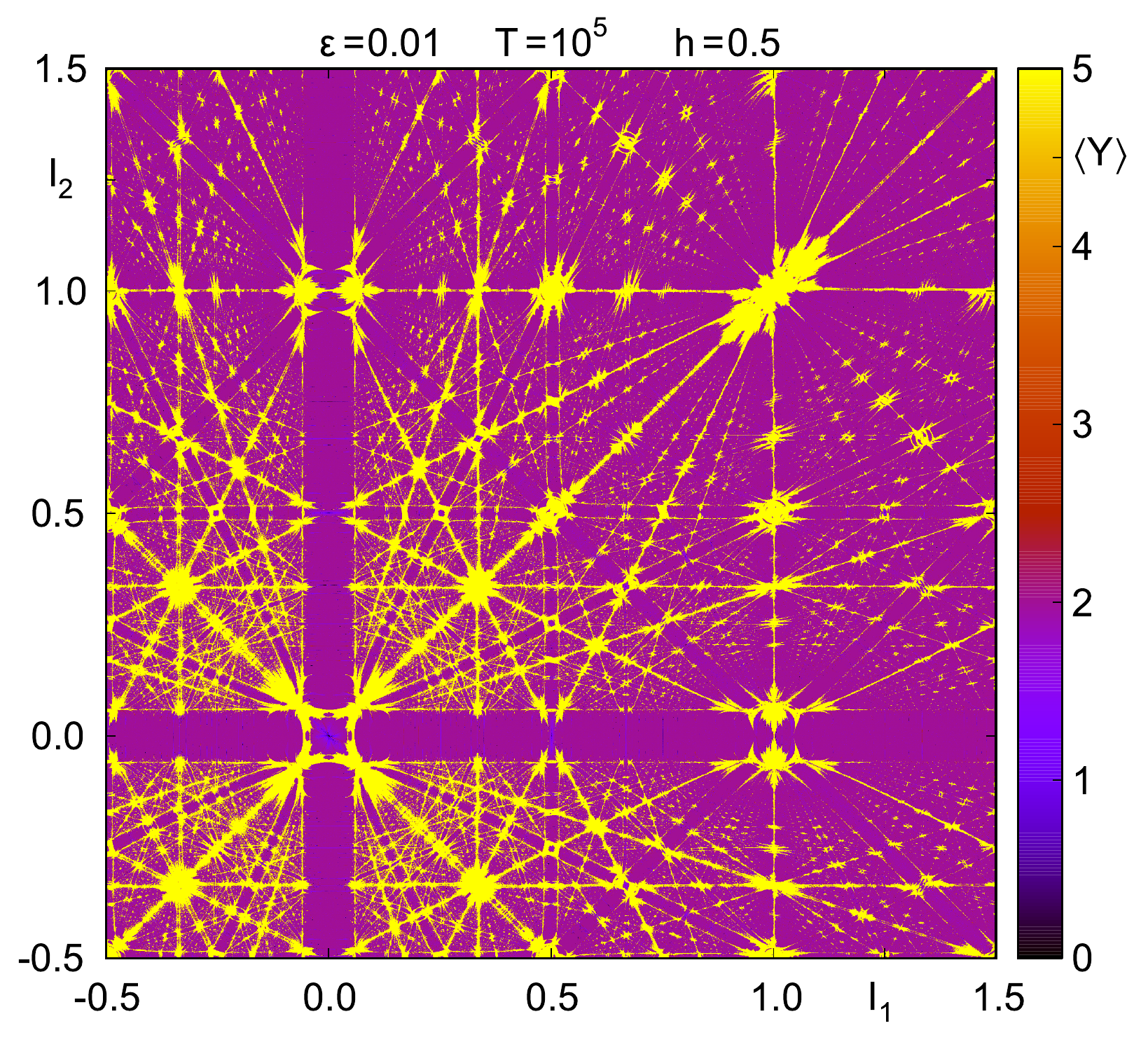}
   \includegraphics[     width=0.499\textwidth]{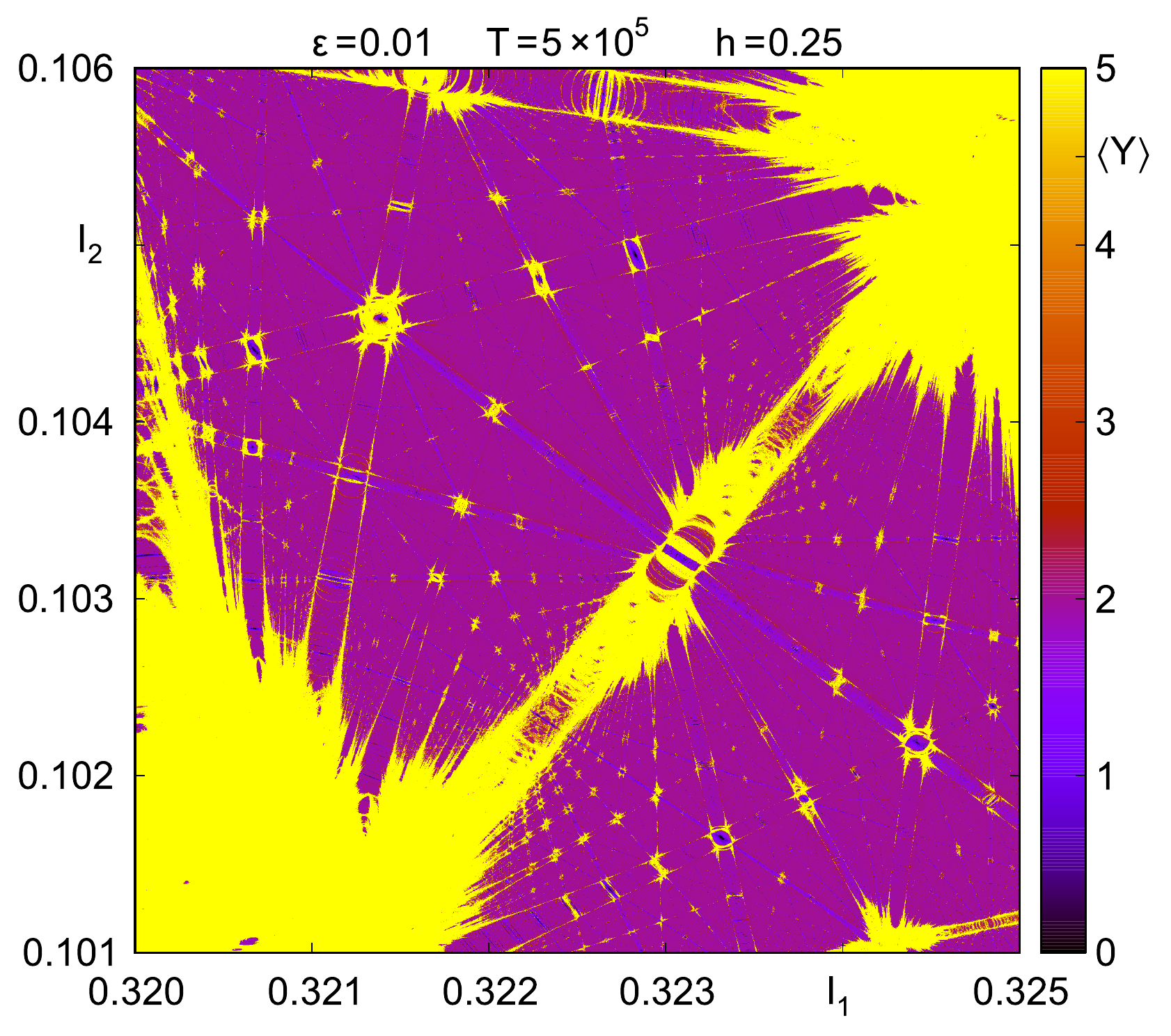}
 }
}
}
\caption{
MEGNO dynamical maps of the model Hamiltonian (Eq.~\ref{eq:eq1}) in the actions 
$(I_1,I_2)$-plane, corresponding to the fundamental frequencies of the 
unperturbed model. Panels are for $\epsilon=0.01$. The left 
panel shows the quasi-global view of the phase space, with a close-up on the right. 
Solutions are colour-coded: 
stable solutions with the MEGNO $\sim 2$, unstable (chaotic) with the MEGNO 
cut-off at 5, and stable periodic orbits are marked with the MEGNO $\sim 0$. 
Original, raw resolution of the maps is $2048 \times 2048$ points. The 
step-size $h$ of the symplectic SABA$_3$ integrator and total integration 
time are labeled.
}
\label{fig:fig1}
\end{figure*}

\section{Hamiltonian model of the Arnold Web}\label{sec:arnold}
We illustrate programming concepts behind the \mechanic{} environment with the dynamical
system derived by \cite{Froeschle2000}. The system is written in Hamiltonian form of
\begin{equation}
{\cal H} = {\cal H}_0(I_1,I_2,I_3) + \epsilon V(\phi_1,\phi_2,\phi_3),
\label{eq:eq1}
\end{equation}
where the Hamiltonian terms are
\begin{eqnarray}
 {\cal H}_{0} &=& \frac{1}{2} I_1^2 + \frac{1}{2} I_2^2 + I_3,  \\ 
 {\cal V}     &=& \frac{1}{\cos\phi_1 + \cos\phi_2 + \cos\phi_3 +4}.
\end{eqnarray}
Actions $I_1,I_2,I_3 \in {\mathbb R}$ and angles $\phi_1, \phi_2, \phi_3 \in 
{\mathbb T}$ are canonically conjugated variables, and $\epsilon$ is a 
parameter that measures the perturbation strength. Indeed, if $\epsilon=0$, 
the equations of motion of Hamiltonian ${\cal H}_0$ are trivially 
integrable. Because angles are cyclic in ${\cal H}_0$, actions 
$I_1,I_2,I_3$ are constant; then the angles are linear functions of time 
$\phi_i = f_i t + \phi_i(0)$, where
\[
f_i = \frac{\partial {\cal H}}{\partial I_i} \equiv
           \frac{\partial {\cal H}_0}{\partial I_i}, \quad i=1,2,3.
\]
The motions generated by the integrable Hamiltonian are confined to 
invariant tori composed of quasi-periodic solutions having the fundamental 
frequencies $f_1=I_1$, $f_2=I_2$, $f_3=1$. With the 
perturbation term $\epsilon\neq0$, the full dynamics are non-integrable. 
According with the Kolmogorov-Arnold-Moser theorem \citep[KAM,][]
{Arnold1978}, the quasi-periodic solutions persist in some volume of the 
phase space, provided that certain non-degeneracy conditions are fulfilled 
and the unperturbed tori are sufficiently non-resonant:
\[
k_1 f_1 + k_2 f_2 + k_3 f_3 \neq 0, \quad k_1, k_2, k_3 \in {\mathbb Z}.
\]
The KAM theorem does not apply in a neighborhood of the resonances, which 
are represented as lines, up to the distance of the order of 
$\sqrt{\epsilon} \exp(-\norm{\v{k}})$, where $\norm{\v{k}}$ is the order of the 
resonance. In that zone, called {\em the Arnold web}, the dynamics are 
extremely complex. 
\begin{figure*}
\centerline{
\vbox{
 \hbox{
   \includegraphics[     width=0.499\textwidth]{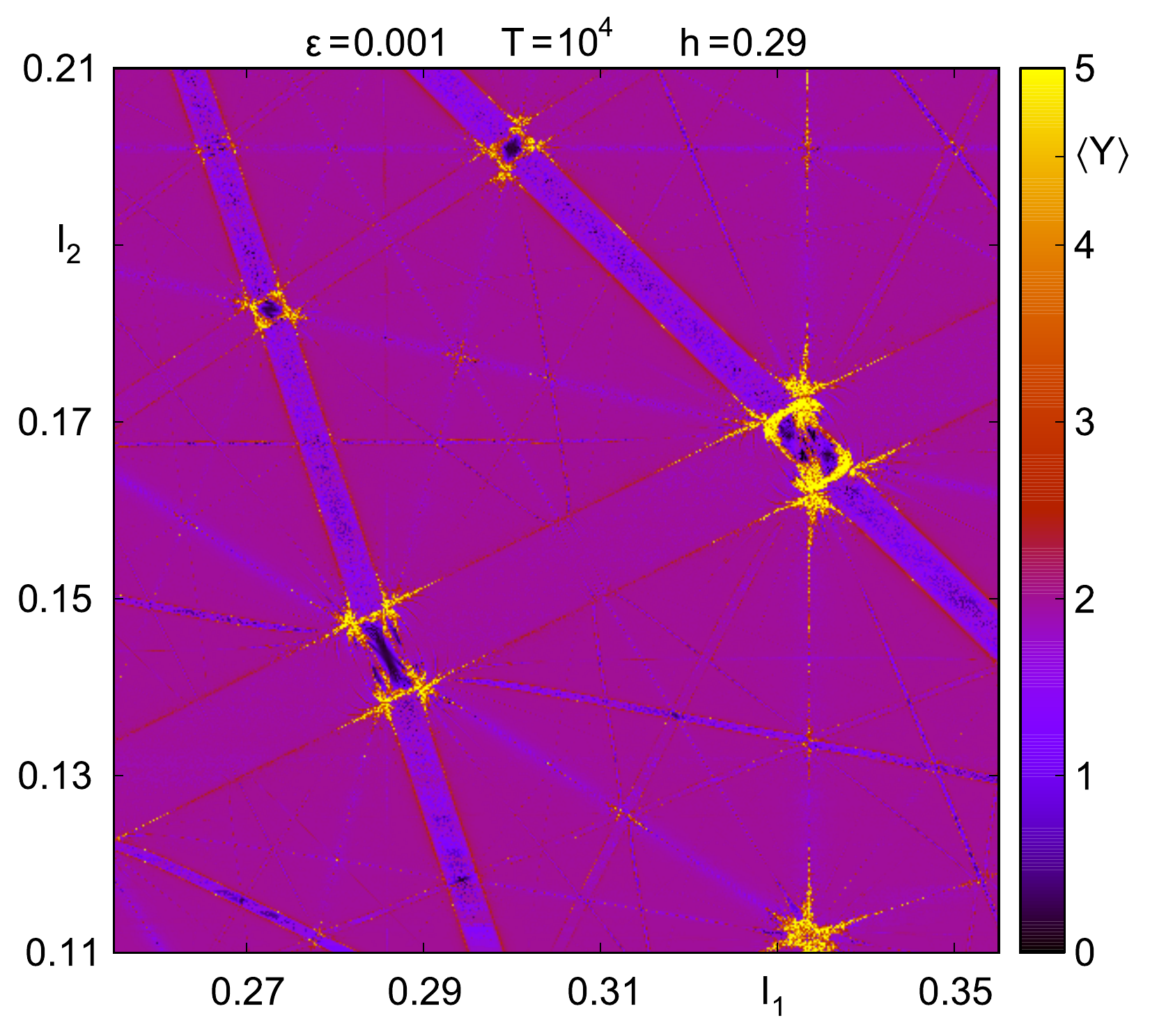}
   \includegraphics[     width=0.499\textwidth]{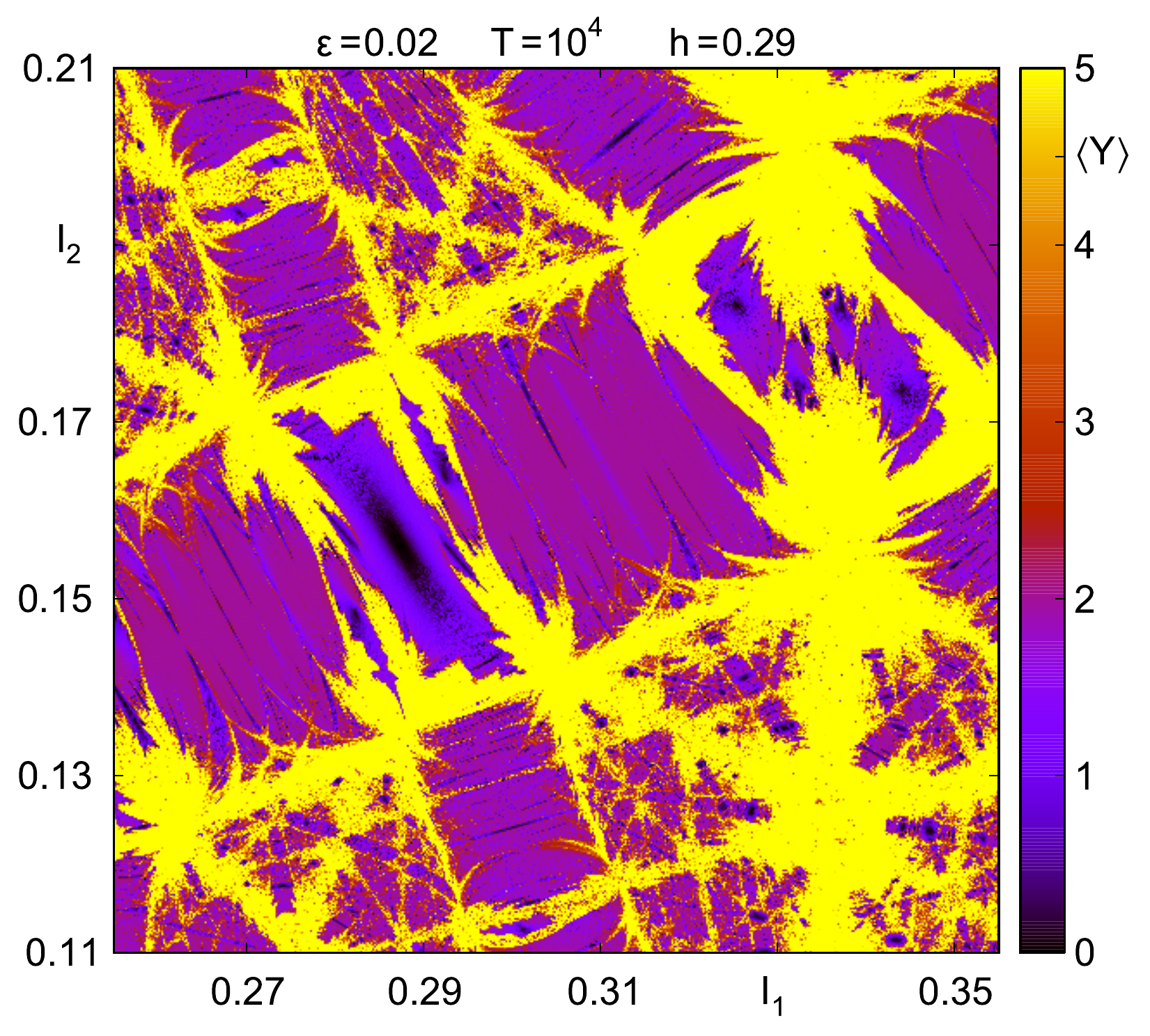}
 }
 \hbox{
   \includegraphics[     width=0.499\textwidth]{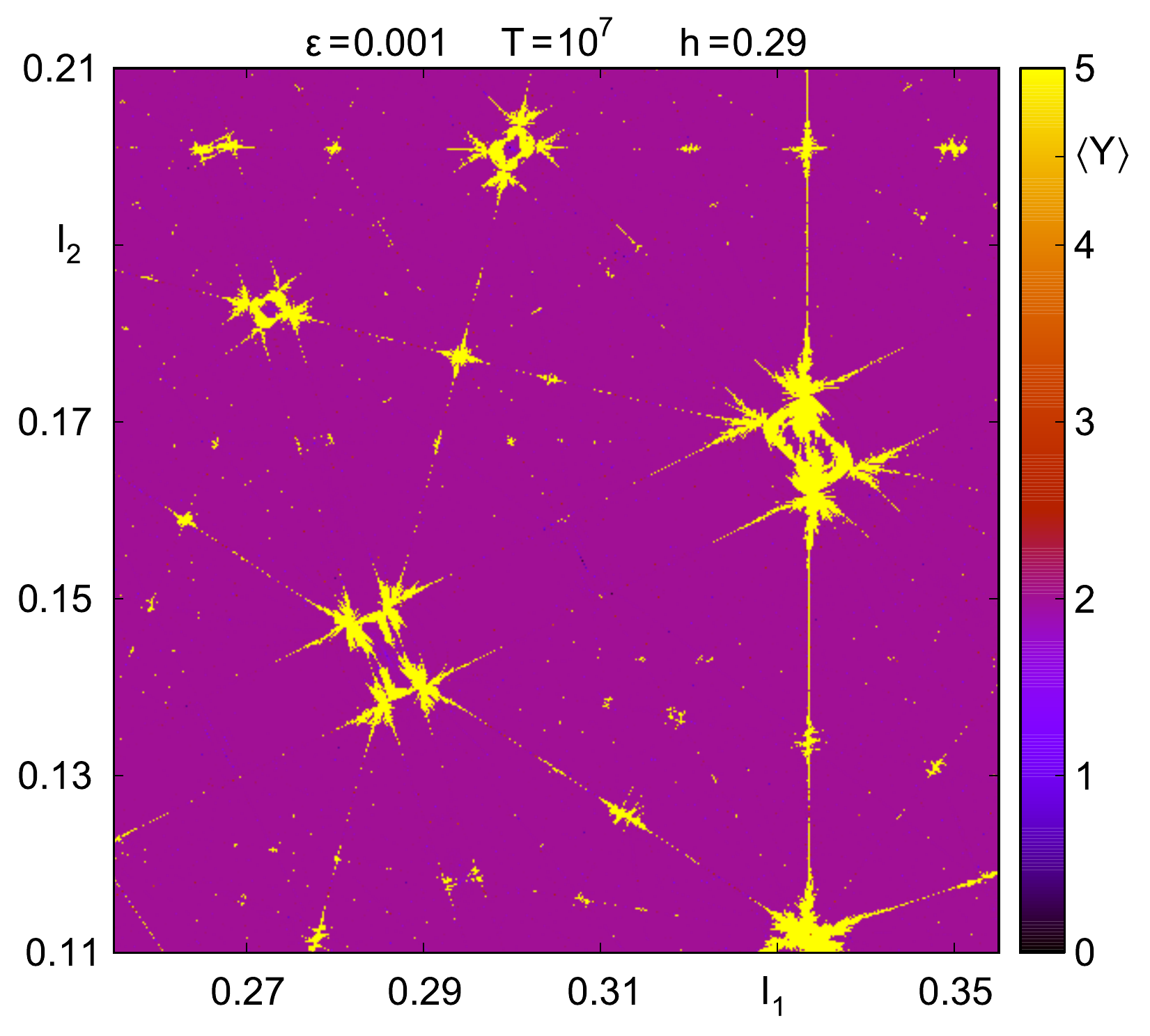}
   \includegraphics[     width=0.499\textwidth]{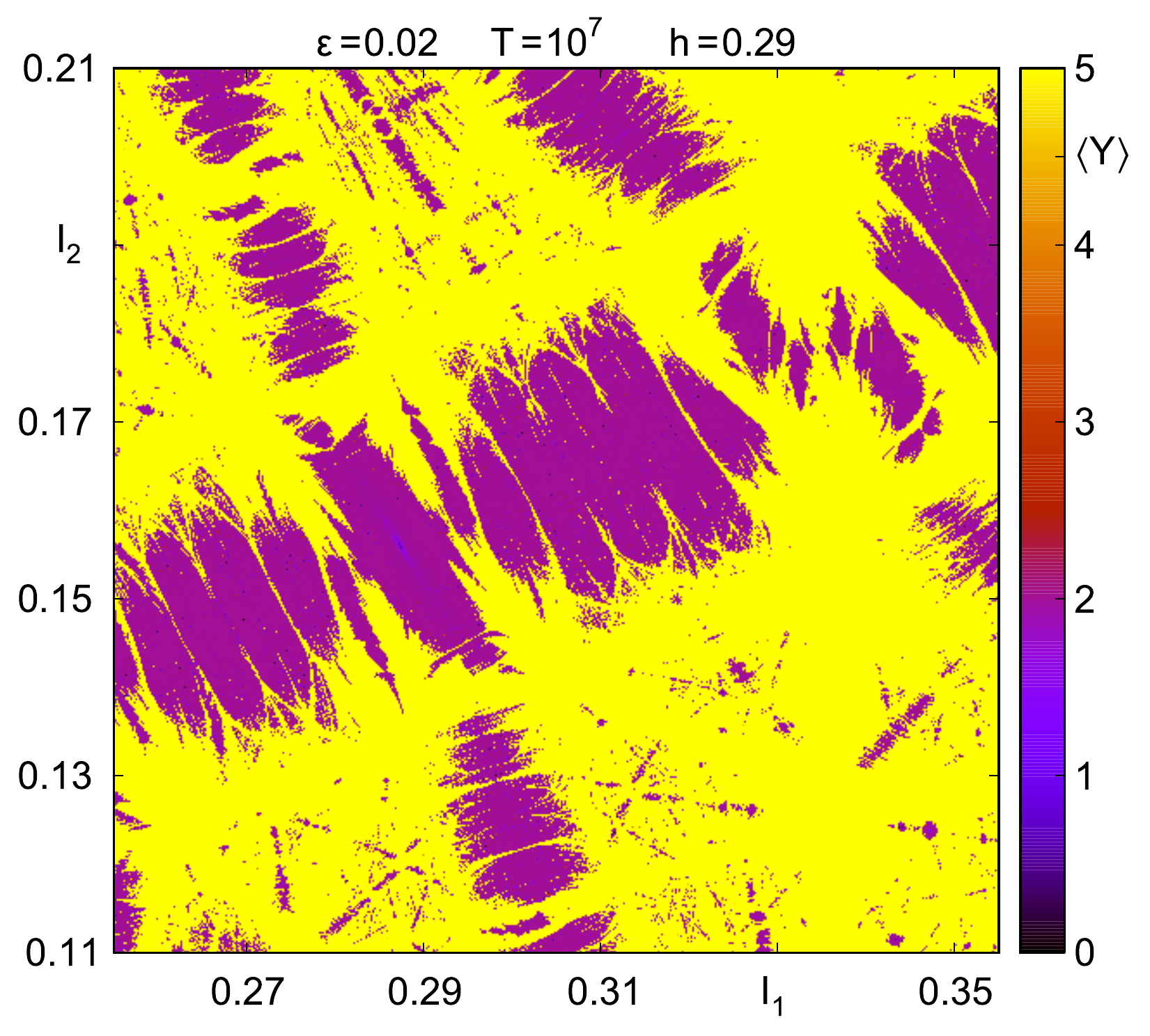}
 }
 \hbox{
   \includegraphics[     width=0.499\textwidth]{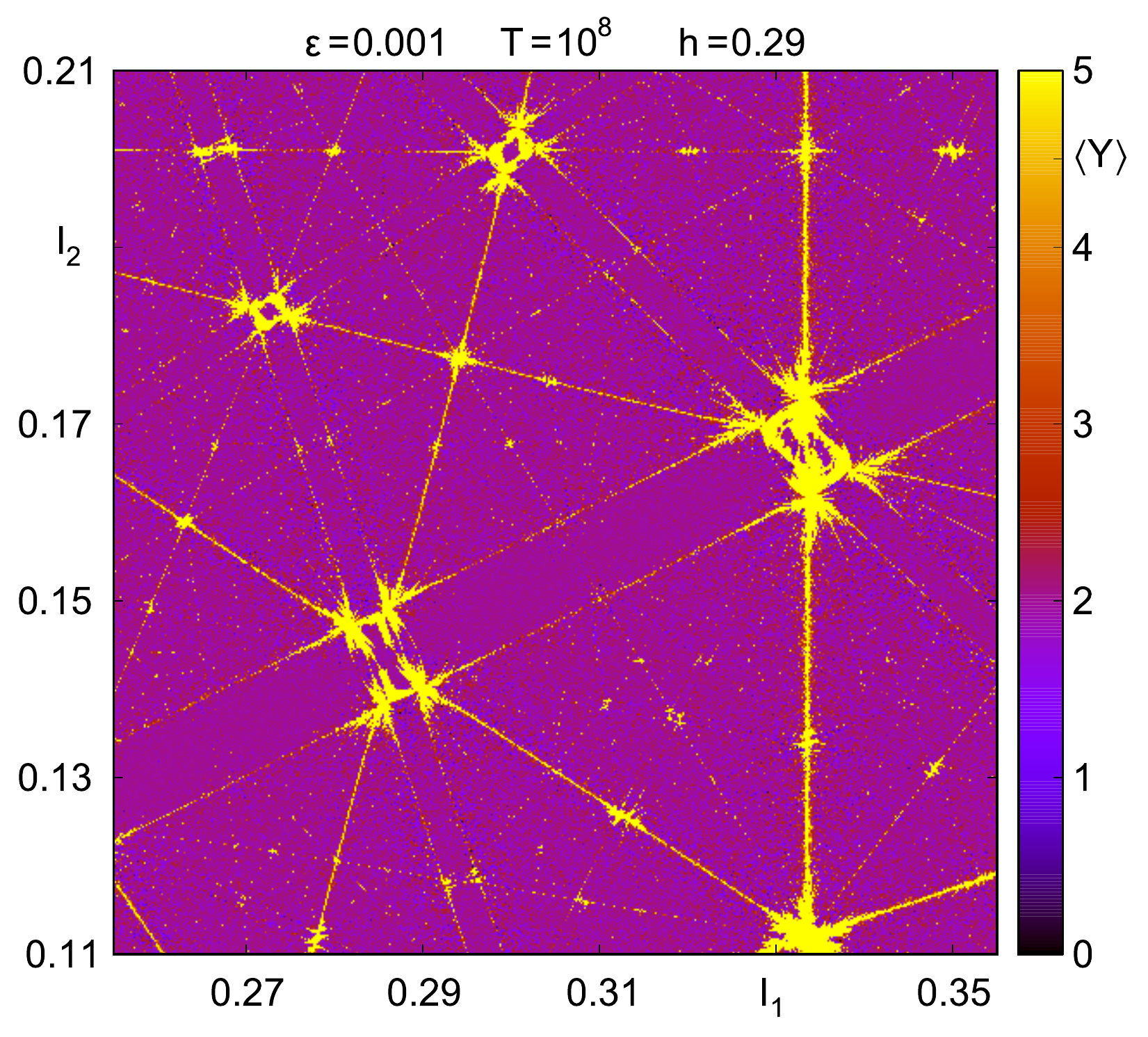}
   \includegraphics[     width=0.499\textwidth]{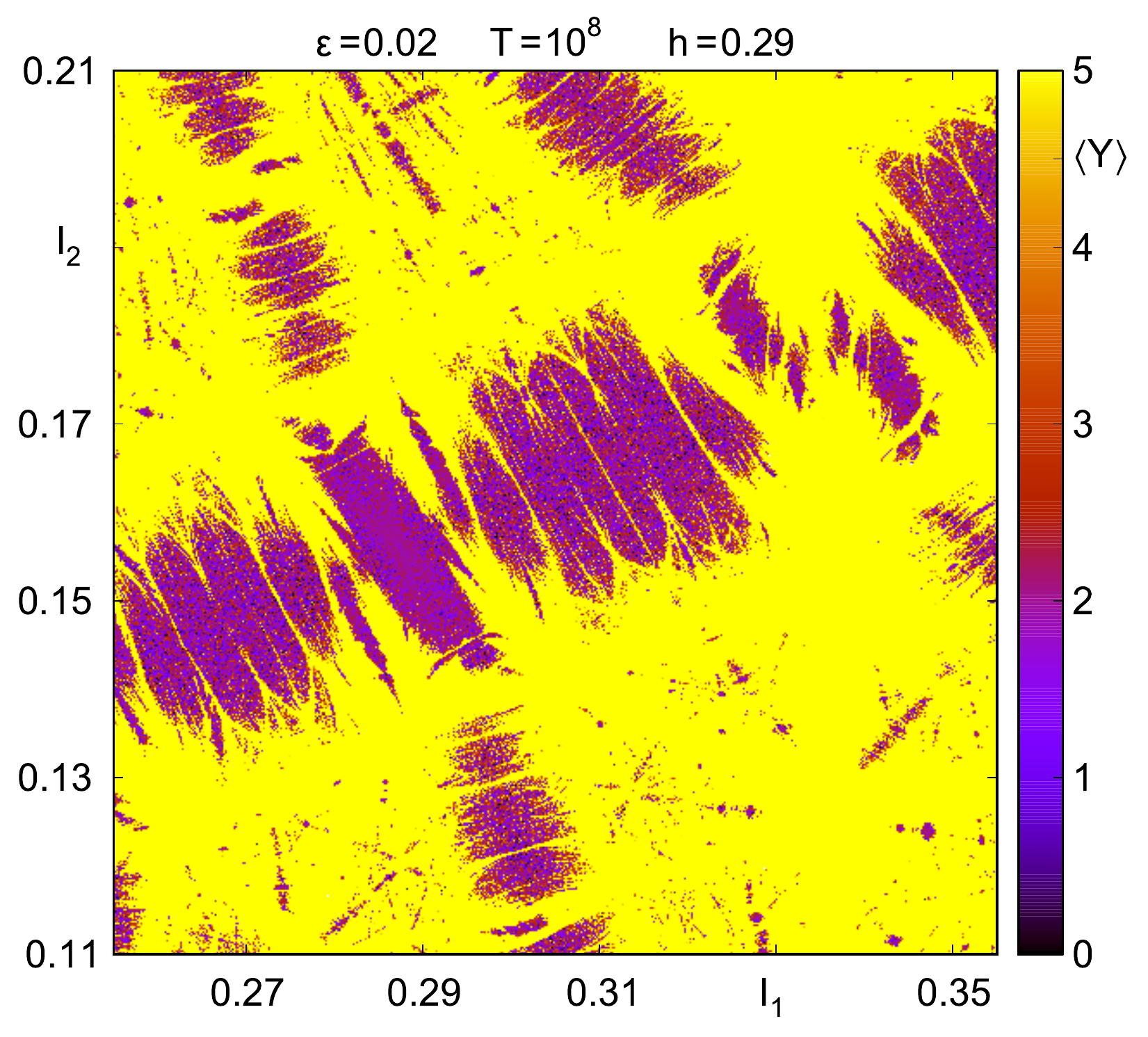}
 }
}
}
\caption{
The Arnold diffusion shown with the MEGNO
time-snapshots in a close-up region of the model Hamiltonian (Eq.~\ref{eq:eq1}) near the
resonant line $I_1 = 2I_2$. Left panels are for $\epsilon=0.001$, which corresponds to the
stable Nekhoroshev regime and the right for $\epsilon=0.02$, corresponding to the transition
region between Nekhoroshev and Chirikov regimes \citep{Lega2003}.
The stepsize of the symplectic SABA$_3$ and the total integration time are labeled. The
raw resolution of the maps is $512\times512$ points.
}
\label{fig:fig2}
\end{figure*}

\begin{figure*}
\centerline{
\vbox{
 \hbox{
   \includegraphics[     width=0.499\textwidth]{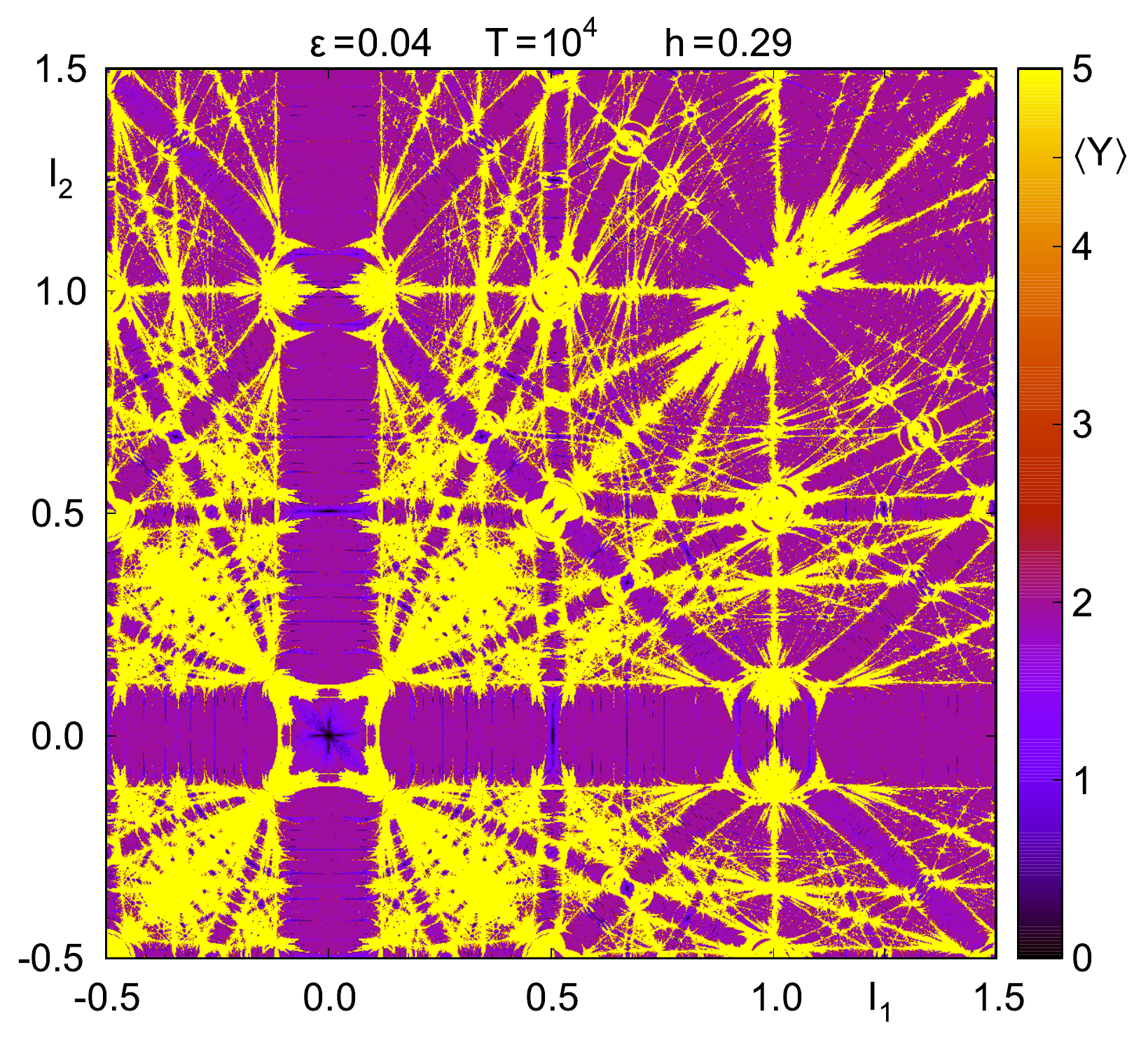}
   \includegraphics[     width=0.499\textwidth]{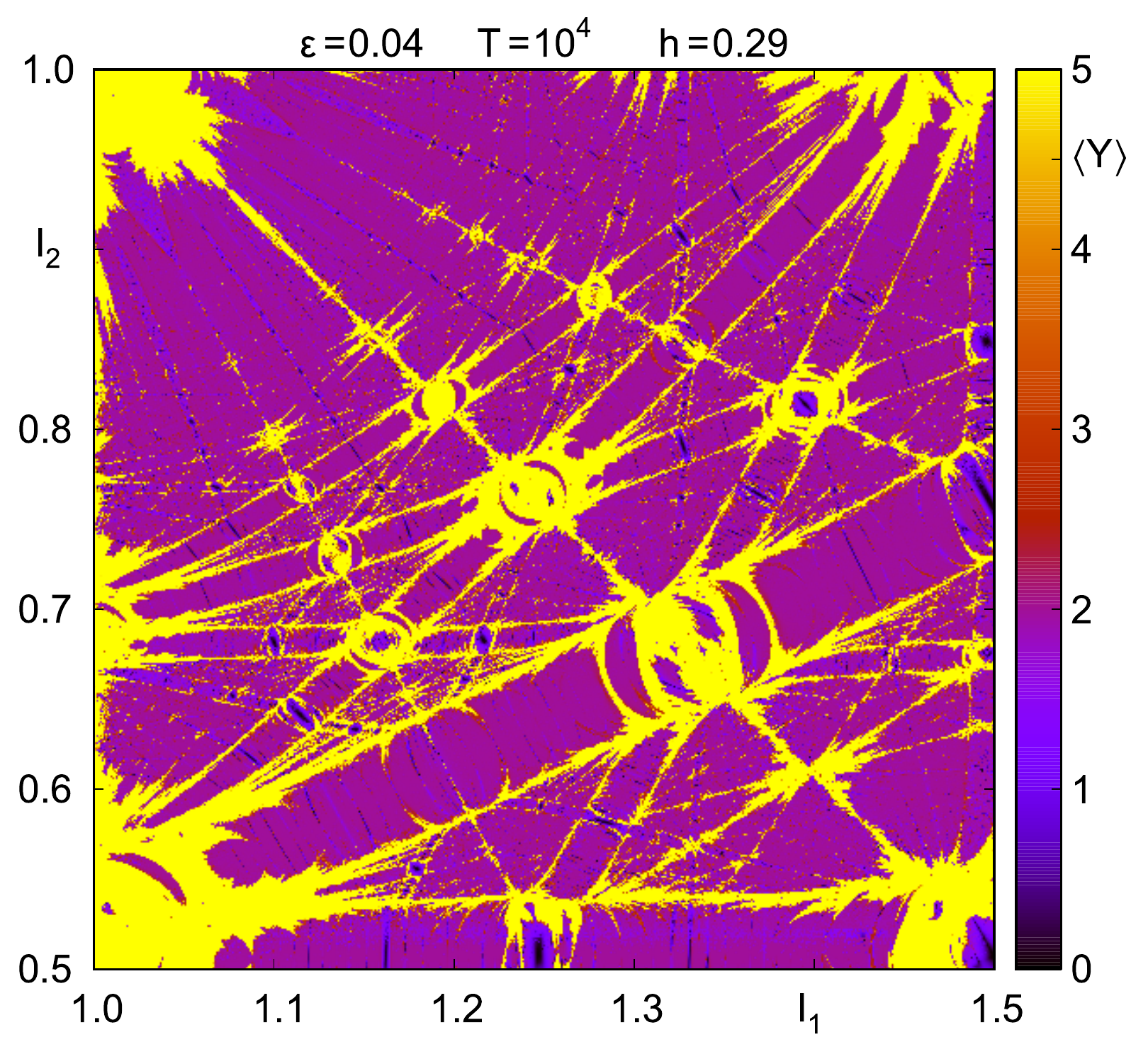}
 }
 \hbox{
   \includegraphics[     width=0.499\textwidth]{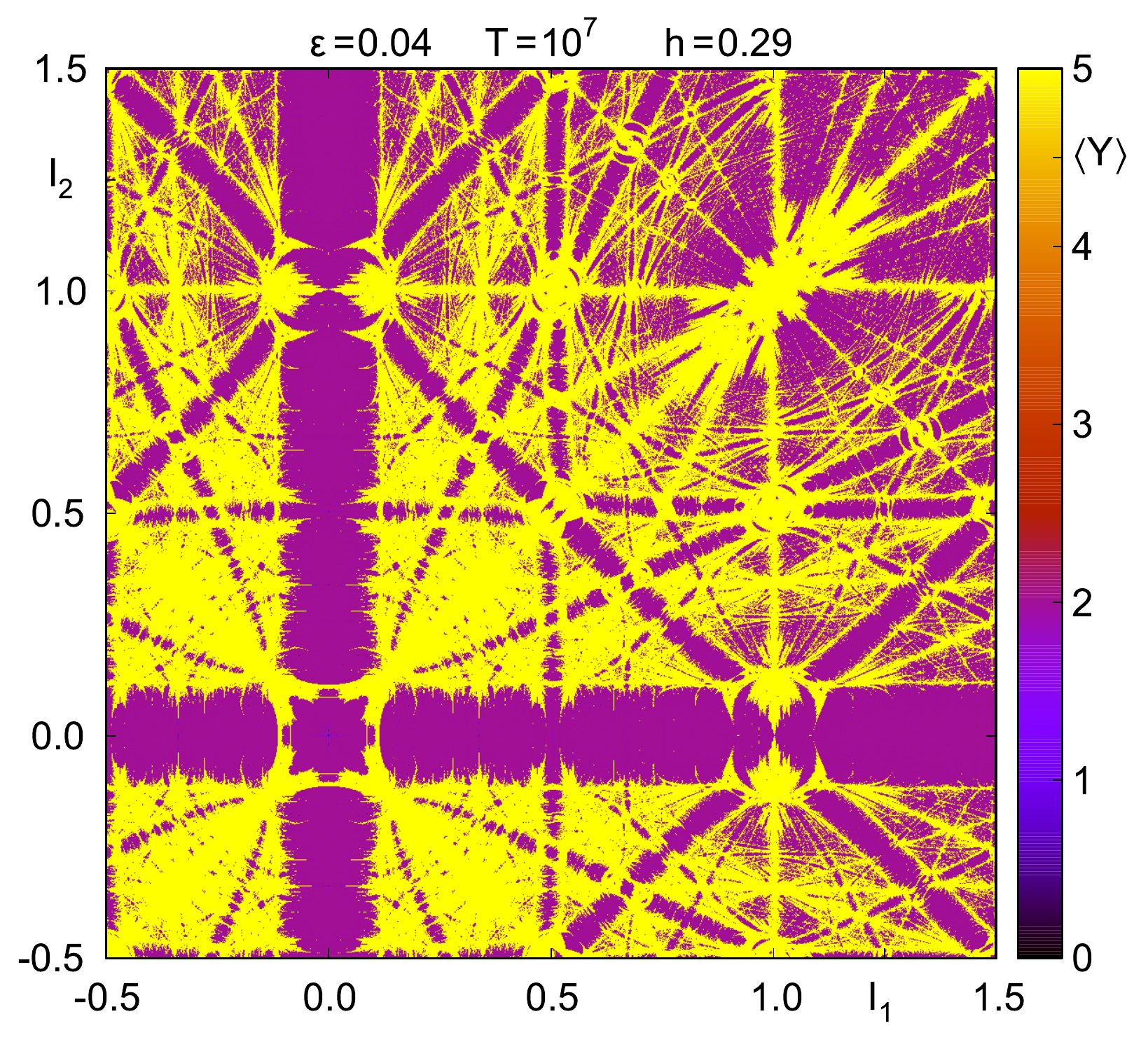}
   \includegraphics[     width=0.499\textwidth]{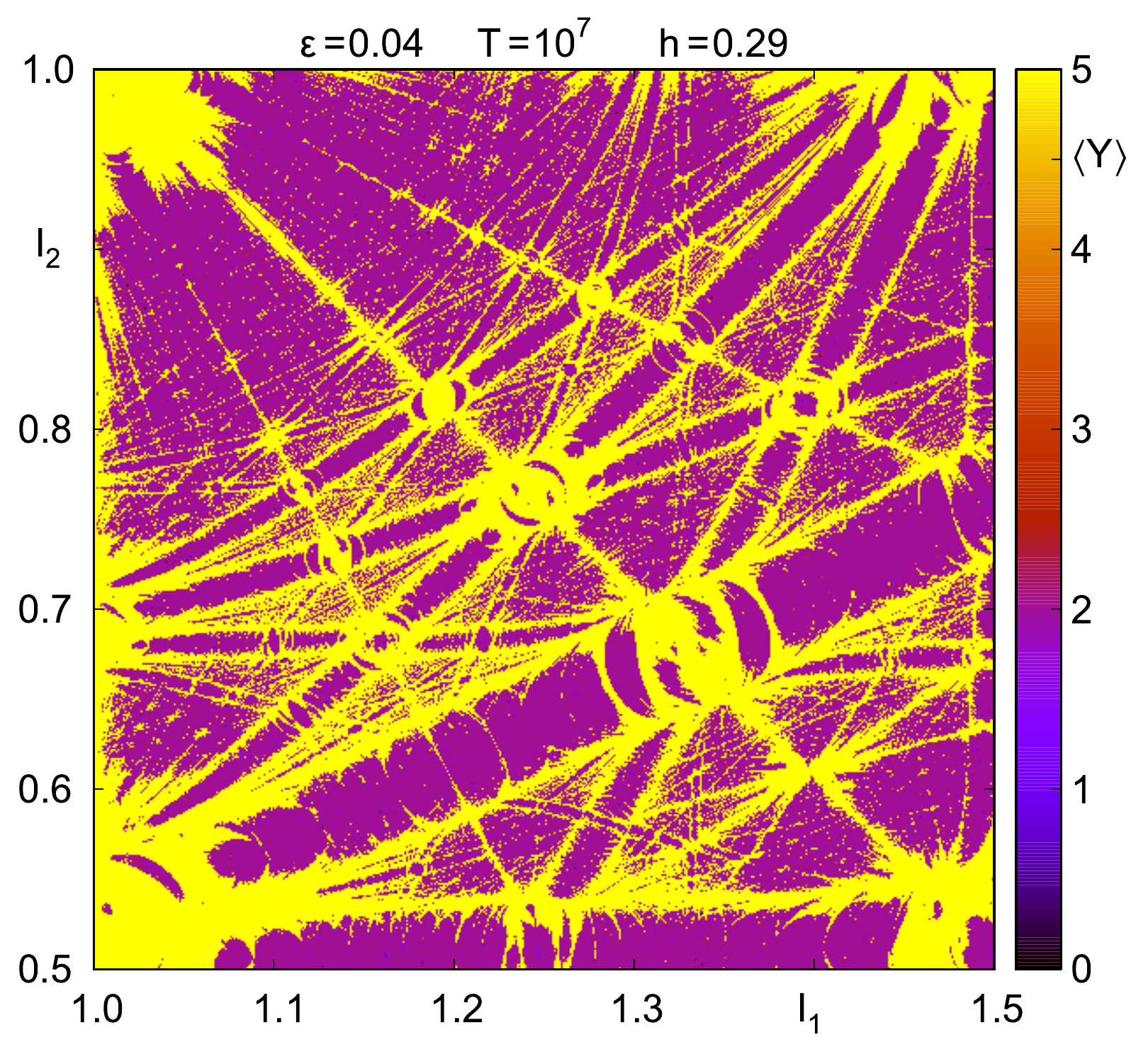}
 }
 \hbox{
   \includegraphics[     width=0.499\textwidth]{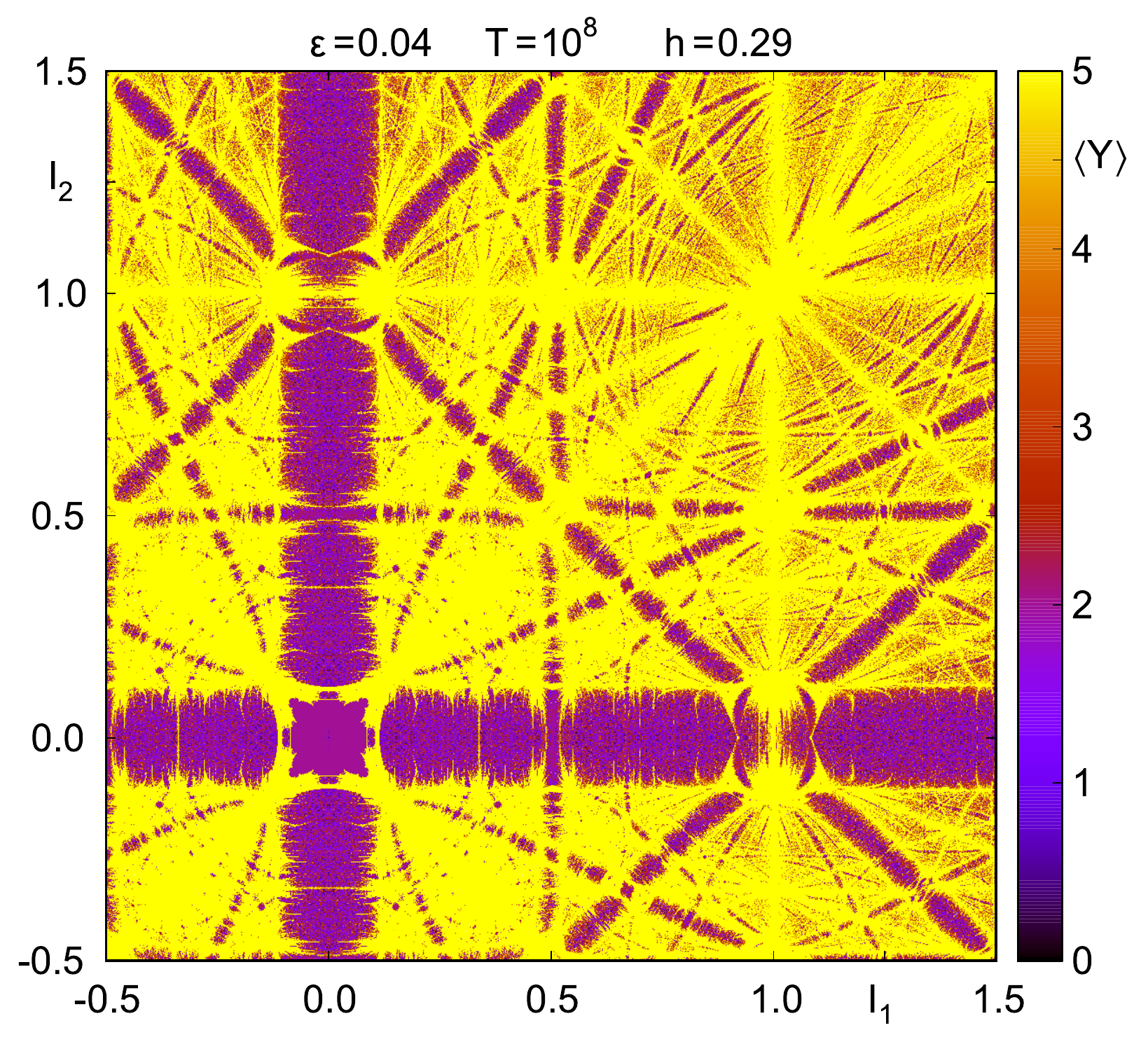}
   \includegraphics[     width=0.499\textwidth]{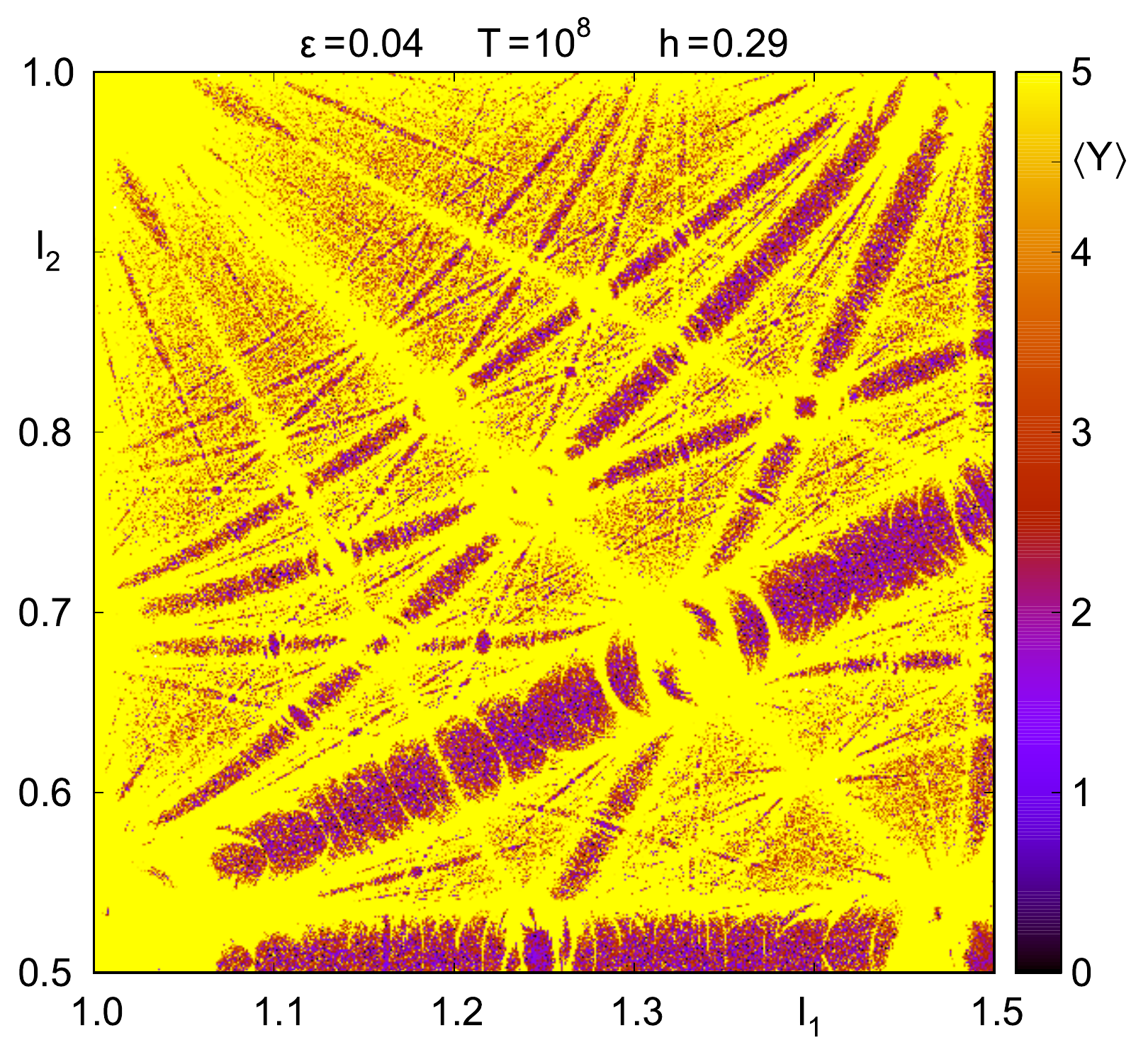}
 }
}
}
\caption{
The evolution of the model Hamiltionian (Eq.~\ref{eq:eq1}) shown with the MEGNO
time-snapshots. All panels are for $\epsilon=0.04$. The left panels show the global view
in the actions $(I_1,I_2)$-plane. The close-up near the resonant line $I_1 = 2I_2$ is
shown on the right.
The stepsize of the symplectic SABA$_3$ and the total integration time are labeled. The
raw resolution of the global maps is $1024\times1024$ points, and the close-ups
$512\times512$ points.
}
\label{fig:fig3}
\end{figure*}

Following \cite{Froeschle2000}, we  visualize the structure of the 
Arnold web through applying a concept of dynamical maps. The Arnold web can be 
represented in two-dimensional actions plane, e.g., $(I_1,I_2)$, which 
correspond to fundamental frequencies of the unperturbed system \citep
{Froeschle2000}.  To detect the regular and chaotic motions, 
which are expected in a non-integrable Hamiltonian system, we apply the 
fast indicator MEGNO \citep{cs2000}. Recently, \cite{Mestre2011} showed analytically that 
this numerical tool brings essentially the same information as the Fast 
Lyapunov Indicator used by \cite{Froeschle2000}. They computed dynamical maps 
of the Arnold web for a few representative values of $\epsilon=0.001$, 
$\epsilon =0.01$ and $\epsilon=0.04$, with the resolution of $500 \times 
500$ pixels. 

With the help of \mechanic{}, we attempt to illustrate the Arnold web in 
much larger resolutions revealing very fine details of the phase space. 
They appear due to resonances of large orders. It is only a matter of long 
enough integration time to detect all resonances, but much longer motion 
times than $10^3$ characteristic periods ($\equiv 1$) in the original paper 
are required. The general-purpose integrators, like the Runge-Kutta or 
Bulirsh-Stoer-Gragg schemes are not accurate nor efficient enough for that 
purpose. These methods introduce systematic drift of the energy (or other 
integrals). To avoid such errors, and to solve the variational equations 
required to compute the MEGNO indicator, we applied the symplectic tangent 
map algorithm introduced by \cite{Mikkola1999}. In the past, we used this 
scheme for an efficient and precision computations of MEGNO for multiple 
planetary systems \citep{Gozdziewski2003,Gozdziewski2005,Gozdziewski2008}.

The model Hamiltonian is particularly simple to illustrate the symplectic 
algorithm. It relies on concatenating maps $\Phi_{1,h}(\v{I},\v{\phi})$ and 
$\Phi_{2,h}(\v{I},\v{\phi})$ that solve the equations of motion derived 
from the unperturbed part ${\cal H}_0$, and the perturbation alone, on the 
time interval $[t_0, t_0+h]$. The solutions may be constructed if both 
Hamiltonian terms admit analytical solutions. This is the case. For the 
unperturbed term we have
\[
 \frac{d}{dt}{\v{\phi}} = \frac{\partial {\cal H}_0}{\partial \v{I}}, \quad
 \frac{d}{dt}{\v{I}} = -\frac{\partial {\cal H}_0}{\partial \v{\phi}}.
\]
Then the ``drift'' map $\Phi_{1,h}(\v{I},\v{\phi})$ map is the following:
\[
 {\v{\phi}} = \v{\omega} h + {\v{\phi}}_0, \quad {\v{I}} = {\v{I}}_0,
\]
where $\v{f} = (I_1, I_2, 1)$. The equations of motion generated by 
the perturbation Hamiltonian  $H_1 \equiv \epsilon V(\v{\phi})$ alone are 
also soluble, hence we obtain the ``kick'' map $\Phi_{2,h}(\v{I},\v{\phi})$:
\[
{\v{\phi}} =  {\v{\phi}}_0, \quad {\v{I}} = \v{\Omega} h + {\v{I}}_0, \qquad
 \v{\Omega} = -\epsilon  \frac{\partial {\cal V}(\v{\phi)}}{\partial \v{\phi}}.
\]
A classic leap-frog concatenation of these maps 
\[
\Phi \equiv \Phi_{1,h/2} \odot \Phi_{2,h} \odot  \Phi_{1,h/2},
\]
provides a numerical integrator of the second order with local error ${\cal 
O}(\epsilon h^2)$. However, because we deal with a small perturbation 
parameter, much better performance and accuracy can be obtained by applying 
symplectic schemes invented by \cite{Laskar2001}. The $n$-th order 
integrator, e.g., {\cal SABA}$_n$, have the truncation error ${\cal 
O}(\epsilon^2 h^2 + \epsilon h^n)$. For small $\epsilon$ it behaves like a 
higher-order scheme without introducing negative sub-steps. In our 
computations, we used the {\cal SABA}$_3$ scheme. We found that it provides 
an optimal performance, CPU-overhead vs. the relative error of the energy.

To compute the MEGNO, we must solve the variational equations to the equations 
of motion. We use the algorithm described in \citep{Gozdziewski2008}. The 
tangent map approach \citep{Mikkola1999}  requires to differentiate the 
``drift'' and ``kick'' maps. This step is straightforward. The variations 
are propagated within the same symplectic scheme, as the equations of 
motion. Having the variational vector $\v{\delta}$ computed at  discrete 
times, we find temporal $y$ and mean $Y$ values of the MEGNO indicator at 
the $j$-th integrator step ($j=1,2,\ldots$) in the form of \citep
{Cincotta2003,Gozdziewski2008}:
\begin{eqnarray*}
Y(j) &=& \frac{(j-1) Y(j-1) + y(j)}{j}, \\
y(j) &=& \frac{j-1}{j} y(j-1) + 2\ln \left(\frac{\delta_j}{\delta_{j-1}}\right),
\end{eqnarray*}
with initial conditions $y(0)=0$, $Y(0)=0$, $\delta = \norm{\v{\delta}}$. 
Following \cite{Cincotta2003}, the MEGNO map obtained in this way tends 
asymptotically to
\[
 Y(j) = a h j + b,
\]
where $a=0,b \sim 2$ for quasi-periodic orbits, $a=b=0$ for stable, periodic 
orbit, and $a=(1/2)\sigma, b=0$ for chaotic orbit with the maximal Lyapunov 
exponent $\sigma$. 

Thanks to the linearity of the 
tangent MEGNO map, the variational vector can be normalized, if its value 
grows too large for chaotic orbits. In practice, we stop the integration if 
MEGNO reaches a given limit ($Y=5$ in this particular case).

\section{The results}\label{sec:results}

We conducted simulations of MEGNO maps in parallel using the \mechanic{} framework up
to 2048 CPUs installed at the \code{Reef}, \code{Cane} and \code{Chimera} clusters (Pozna\'n
Supercomputing Centre, PCSS). Simulations were performed using the master--worker
communication mode. The resolution of the maps varied between $512\times512$ up
to $2048\times2048$ points. The actions $(I_1,I_2)$-plane has been regularly spaced
according to the run-time configuration and the task coordinates on the task assignment
grid. For each initial condition, different initial variational vector has been choosen. 
The other initial conditions were: $I_3(0) = 1$ and $\phi_1(0) = \phi_2(0) = \phi_3(0) = 0$. 
Technical details of the module implementation are given in the Appendix.

The results for model Hamiltonian (Eq.~\ref{eq:eq1}) are illustrated on the
Fig.~\ref{fig:fig1}. The left panel shows the quasi-global view of the $(I_1,I_2)$-plane
for the $\epsilon=0.01$ and $T=10^5$. The resonances appear as straight lines. 
The yellow (light-gray) colour encodes chaotic orbits, and the purple (dark-gray) colour denotes $\langle Y\rangle\sim2$
of stable, quasi-periodic solutions. According to \cite{Froeschle2000}, the dynamics is
governed by the Nekhoroshev regime, when the most of invariant tori of the perturbed
system exist. However, as shown in \cite{Lega2003}, diffusing orbits exists along the resonances. 
This leads to significant drifts in the actions space
\citep{Guzzo2004,Froeschle2005,Froeschle2006}.
This phenomenon, called the Arnold diffusion is closely related to the stability of the
system, and was suggested by \cite{Arnold1964} in the three-body problem. 

To illustrate the Arnold diffusion, we computed MEGNO time-snapshots over the
time-span $T=10^4-10^8$. The slow diffusion in the stable Nekhoroshev regime
($\epsilon=0.001$) is shown on the left panels of the Fig.~\ref{fig:fig2}.
Most of solutions remains quasi-stable after $T=10^8$. 
However, for large $\epsilon$, the diffusion is not forced along the resonances. 
Instead, the resonances overlap and the diffusing orbits may wander between different
resonances \citep{Lega2003}. This transition region for $\epsilon=0.02$ is illustrated on
the right panels of the Fig.~\ref{fig:fig2}. 
At the critical value, the region of diffusing orbits
replaces the region of invariant tori, and the dynamics is governed by the Chirikov
regime. As estimated by \cite{Froeschle2000}, the critical $\epsilon$ for the Hamiltonian
(Eq.~\ref{eq:eq1}) is $0.04$ (Fig.~\ref{fig:fig3}).
Most of the phase space becomes chaotic after $T=10^8$.

\begin{figure*}
\centerline{
\vbox{
 \hbox{
   \includegraphics[     width=0.499\textwidth]{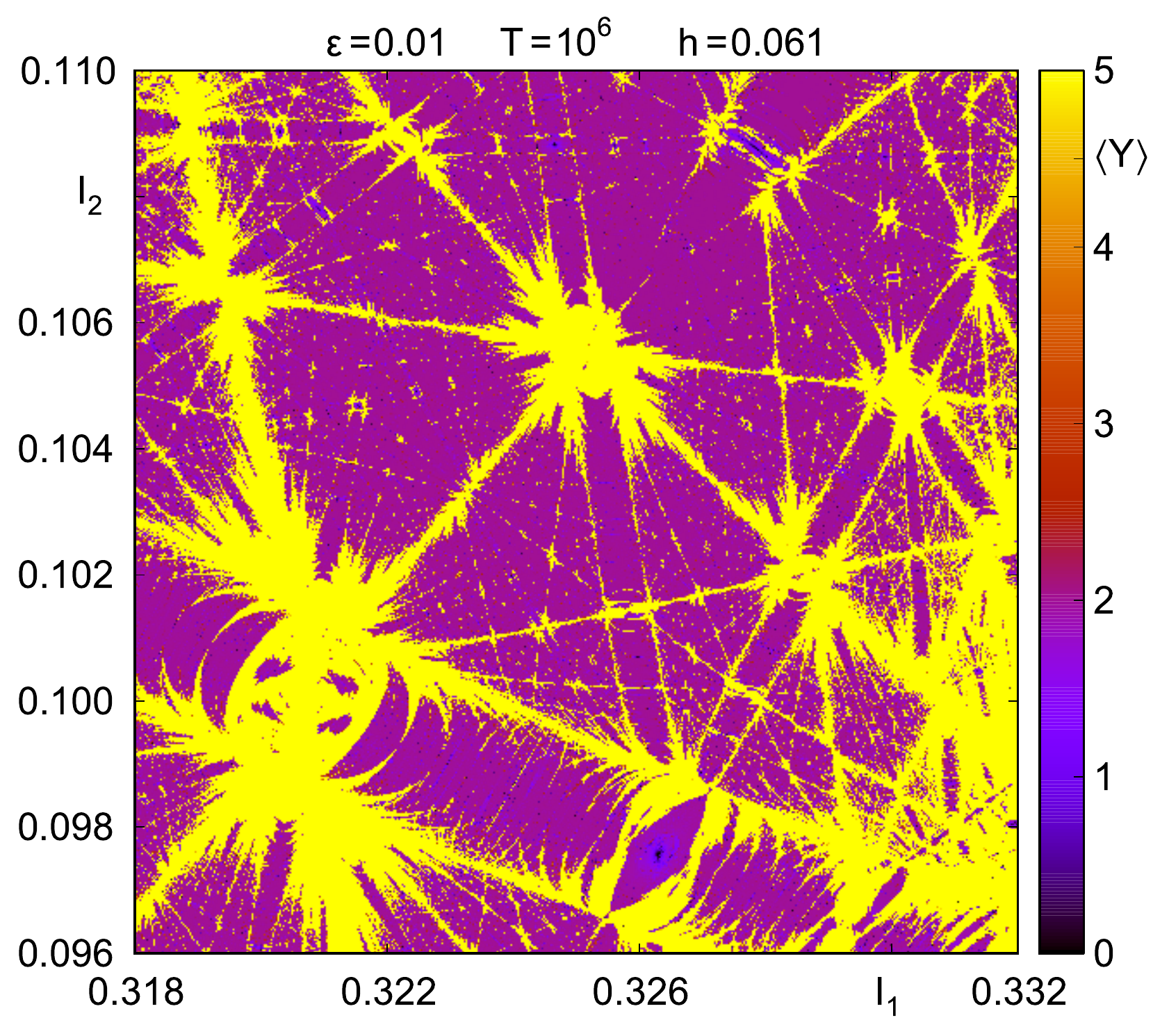}
   \includegraphics[     width=0.499\textwidth]{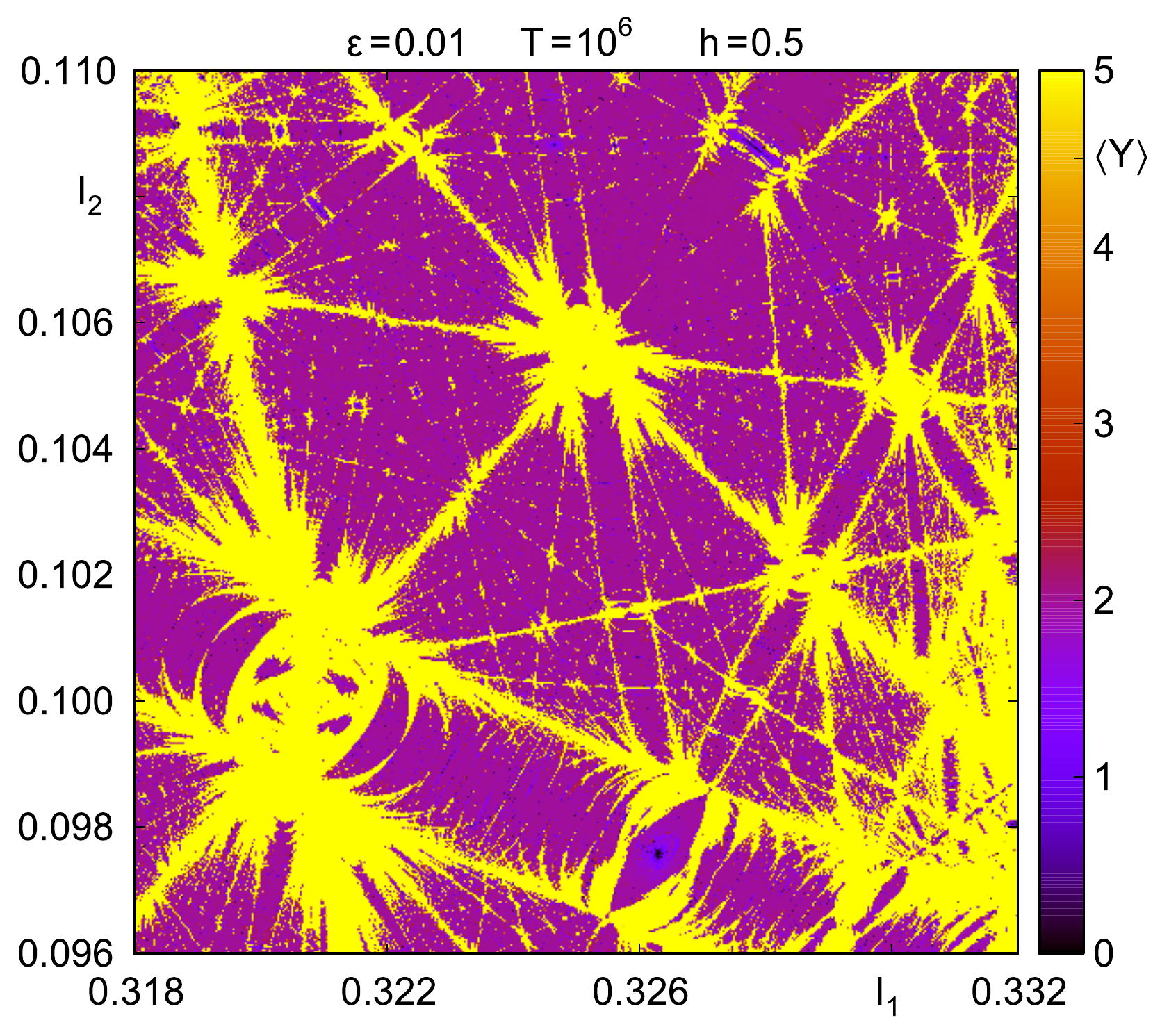}
 }
}
}
\caption{Close-ups of the MEGNO dynamical maps for $\epsilon=0.01$ and $T=10^6$. The left panel shows
the result computed with the step size $\sim0.061$ and the right panel for the step size
$0.5$ of the SABA$_3$ integrator.
}
\label{fig:fig4}
\end{figure*}

Computations were performed with different step-sizes, 
$h=0.01$ up to $h=0.5 (\sqrt{5}-1) \sim 0.29$, and $h=0.5$. 
We found that even 
such large integration steps do not introduce artificial (numerical) 
resonances. As an example, we selected a very small region close to the 
centre of the global map (the left panel in Fig.~\ref{fig:fig1}), and 
we computed MEGNO maps with different step sizes over $T=10^6$ using SABA$_3$ and SABA$_4$
integrators. The results for $h\sim0.06$ and $h=0.5$ are shown on the Fig.~\ref{fig:fig4}.
Both the close-ups reveal very fine 
details of the phase space, and cannot be distinguished each from the 
other. This test assured us, 
that relatively large step-sizes of the order 
of $h \sim 10^{-1}$ are still safe, and the results are reliable.
The relative energy error was preserved up to
$10^{-8} - 10^{-9}$ over the total integration times up to $T \sim 10^8$.

Depending on the characteristic period $T$, the resolution of the maps and 
the number of CPU cores involved, the total simulation time of a single map took up to 72 hours.

\section{Summary} 
We present the \mechanic{} code, an open-source MPI/HDF5 code framework. It is designed to become a
helper tool for conducting massive numerical computations that rely on testing huge
volume of inital conditions, such as studies of long-term orbital
evolution and stability of planetary systems with Monte Carlo methods or dynamical maps, 
as well as modelling observations with evolutionary
algorithms. Based on the {\em core--module}
approach, the framework reduces the development effort of a scientific application, by
providing skeleton code for common technical operations, such as data and configuration
management, as well as task distribution on the parallel computing environments. 
Within the framework, the user's application may be reduced to a module form containing
the numerical algorithm with the required setup and storage specifications.
Unlike the existing task management software, such as \code{HTCondor} or \code{Workqueue}, our framework is based on the
unified data storage approach, that is built on top of the HDF5 library. This reduces the post-processing
effort and makes it possible to use results of the simulation in numerous
external applications written in different programming languages, such as Python. The
framework supports multidimensional datasets of all basic datatypes with attributes. The
communication and storage layers are hidden to the end user, so that, no knowledge on the
parallel programming is required to use the framework.

The \mechanic{} has been already extensively tested in the dynamical studies of
the Kepler-11 \citep{kepler-11} and $\nu$Oct systems \citep{nuoct}. They validated the framework as a
helper tool for astronomical research. The numerical computations have been conducted up
to 2048 CPUs on the \code{Reef}, \code{Cane} and the SGI UV \code{Chimera} supercomputer located at the Pozna\'n Supercomputing Centre. 
In this paper, we have shown the usage of the framework with the simple dynamical map
algorithm applied to the Hamiltonian model of the Arnold Web. The source code of the framework 
is shipped with template modules, that show every aspect of the API. This includes using
predefined initial conditions list for the input, programmatic generation of initial
conditions as well as simple genetic algorithm implementation. The code is BSD-licensed
and available at the project page. 

\section*{Acknowledgements} We would like to thank Kacper Kowalik for initial
collaboration on the project and helpful discussions regarding the framework code, and
Tobias Hinse for testing the \mechanic{} prototype.

This project is supported by the Polish Ministry of Science and Higher Education
through the grant N/N203/402739. This work is conducted within the POWIEW project of
the European Regional Development Fund in Innovative Economy Programme
POIG.02.03.00-00-018/08.

\bibliographystyle{elsarticle-harv}
\bibliography{ms}

\appendix

\section{Detailed code listings for the Arnold Web module}
A \mechanic{} module is a C-interoperable code compiled to a shared library. It consists
of a set of API functions that correspond to the well-known user's application flow. 
The \code{Init()} hook (Listing~\ref{lst:init-hook}) is used for module-related
initializations, including the number of memory buffers passed between the master and
workers. The setup stage is performed via the information specified in the \code{Setup()}
hook (Listing~\ref{lst:setup-hook}). All configuration options are available during the
run-time, both in the configuration file and the command-line.

The file and memory management is performed according to the specification given in the
\code{Storage()} hook (Listing~\ref{lst:storage-hook}). For the purpose of the Arnold Web
example, we define two data buffers. The {\em result} buffer is dedicated to storing the
master result. Each worker node
allocates the memory according to the specified dimensionality and datatype of the
buffer. In this particular case, the result buffer is of size $1\times1\times10\times2$,
and contains the MEGNO value and relative energy error over a maximum of 10 time-snapshots.
The master node assembles the partial results received from workers into a single,
four-dimensional dataset of {\em texture} type, with the respect to the task location on the grid. 
For example, in a $2048\times2048$ points simulation we will obtain the dataset of size
$2048\times2048\times10\times2$, with third dimension (slice) representing time-snapshots
and the last dimension containing the actual result.
The second buffer, {\em state}, contains
the temporary integration data and is removed after successful completion of all tasks.
Otherwise, it may be used to restart the simulation from the last stored checkpoint. 

The actual computations are performed within the \code{TaskPrepare()} and
\code{TaskProcess()} hooks (Listings \ref{lst:task-prepare-hook} and
\ref{lst:task-process-hook}). We change the initial condition according to the run-time
configuration and the task coordinates on the task assignment grid. This occurs only on
the first time-snapshot indicated by the task checkpoint id \code{cid = 0}. The checkpoint
id increments while the \code{TaskProcess()} returns \code{TASK\_CHECKPOINT}. At each
task checkpoint we change the integration time to obtain snapshots with the power-based
intervals, starting with the minimum characteristic period required by the MEGNO,
$T=10^4$. The simulation is continued with the initial condition that is read from the {\em state}
buffer. The result is returned to the master. After the last snapshot has been completed,
the task is finalized, and the worker node takes the next one, if available.

\begin{lstlisting}[caption={\mcap{The module initialization.} The \code{banks\_per\_task}
setting of the \code{Init()} hook specifies the number of memory buffers available for the
each task.
},label={lst:init-hook},firstline=15,lastline=28]
#include "mechanic.h"
#include <math.h>

#define MAX_SNAPSHOTS 10

int saba3(double *state, double step, 
    double tstart, double tend, double eps, 
    double *megno, double *err);

/* Implements Init() */
int Init(init *i) {
  i->banks_per_task = 2;
  return SUCCESS;
}
\end{lstlisting}

\begin{lstlisting}[caption={\mcap{The \code{Setup()} hook.} Each configuration option is
defined within the module \code{space}. All fields are mandatory. The \code{name} and
\code{shortName} define the long and short option name respectively. The default value
\code{value} is used, when the option has not been overridden in the configuration file
nor the command-line. Supported types: \code{C\_INT}, \code{C\_LONG}, \code{C\_FLOAT},
\code{C\_DOUBLE}, \code{C\_VAL} (boolean) and \code{C\_STRING}. 
},label={lst:setup-hook},firstline=30,lastline=45]
/* Implements Setup() */
int Setup(setup *s) {
  s->options[0] = (options) {
    .space = "aweb", .name = "step", .shortName = '\0',
    .value = '0.5', .type = C_DOUBLE,
    .description = "Integrator step size"
  };
  s->options[1] = (options) {
    .space = "aweb", .name = "eps", .shortName = '\0',
    .value = '0.01', .type = C_DOUBLE,
    .description = "The perturbation parameter"
  };
  s->options[2] = (options) OPTIONS_END;

  return SUCCESS;
}
\end{lstlisting}

\begin{lstlisting}[caption={\mcap{The \code{Storage()} hook.} The data buffers are defined
per task pool \code{p}. They are available during the simulation through the task \code{t}
object. The HDF5 dataset path is configured via the \code{name} field. The maximum
dimensionality of a memory buffer (and the corresponding dataset) is 32
(defined via \code{H5S\_MAX\_RANK}).
},label={lst:storage-hook},linerange={47-48,58-83}]
/* Implements Storage() */
int Storage(pool *p) {
  // Path: /Pools/pool-ID/Tasks/result
  p->task->storage[0].layout = (schema) {
    .name = "result",
    .rank = 4,
    .dims[0] = 1, //y-dim
    .dims[1] = 1, //x-dim
    .dims[2] = MAX_SNAPSHOTS,
    .dims[3] = 2,
    .use_hdf = HDF_NORMAL_STORAGE,
    .storage_type = STORAGE_TEXTURE,
    .datatype = H5T_NATIVE_DOUBLE
  };

  // Temporary storage for the state variables
  p->task->storage[1].layout = (schema) {
    .name = "state",
    .rank = 2,
    .dims[0] = 1,
    .dims[1] = 24,
    .use_hdf = HDF_TEMP_STORAGE,
    .storage_type = STORAGE_LIST,
    .datatype = H5T_NATIVE_DOUBLE
  };

  return SUCCESS;
}
\end{lstlisting}

\begin{lstlisting}[caption={\mcap{The \code{TaskPrepare()} hook.}
The initial condition is prepared according to the current run-time configuration, which
is provided in the pool \code{p} object. The configuration is accessed with the \code{MReadOption}
macro. It takes the current pool object and the option name as arguments and reads the
option value to the specified local variable. 
The {\em state} and {\em result} buffers are prepared at the beginning of the simulation
(the task checkpoint id, \code{cid}, is 0). The initial data from local arrays is written to the
task memory buffers with the \code{MWriteData} macro. It is available then in other API
hooks.
},label={lst:task-prepare-hook},firstline=85,lastline=128]
/* Implements TaskPrepare() */
int TaskPrepare(pool *p, task *t) {
  double state[24];
  double r[1][1][MAX_SNAPSHOTS][2];
  double xmin, xmax, ymin, ymax;
  double stepx, stepy;
  int xres, yres, i;
  
  if (t->cid == 0) {
    MReadOption(p, "xmin", &xmin);
    MReadOption(p, "xmax", &xmax);
    MReadOption(p, "ymin", &ymin);
    MReadOption(p, "ymax", &ymax);

    MReadOption(p, "xres", &xres);
    MReadOption(p, "yres", &yres);

    stepx = (xmax-xmin)/((double)xres);
    stepy = (ymax-ymin)/((double)yres);

    // Initial data
    for (i = 0; i < 24; i++)
      state[i] = 0.0;

    state[3] = xmin + t->location[1]*stepx;
    state[4] = ymin + t->location[0]*stepy;
    state[5] = 1.0;

    // Randomize the tangent vector
    for (i = 6; i < 12; i++)
      state[i] = rand()/(RAND_MAX+1.0);

    MWriteData(t, "state", &state[0]);

    // Prepare the result buffer
    for (i = 0; i < MAX_SNAPSHOTS; i++) {
      r[0][0][i][0] = 0.0;
      r[0][0][i][1] = 0.0;
    }
    MWriteData(t, "result", &r[0][0][0][0]);
  }

  return SUCCESS;
}
\end{lstlisting}

\begin{lstlisting}[caption={\mcap{The \code{TaskProcess()} hook.} Before the
computations, we read the initial condition from the {\em state} buffer to the local
buffer with the
\code{MReadData} macro. The data is passed then to the numerical integrator, among with the current time
interval. After the computations, the result is written to the buffer, and returned to the
master node. If the \code{TaskProcess()} hook returns \code{TASK\_CHECKPOINT}, the worker
continues evaluation of the current task with the new time interval based on the task
checkpoint id. The \code{TASK\_FINALIZE} indicates successfull evaluation of the task and
the worker node takes the next one, if available. In case of any error, the hook may
return error code, as specified in the API documentation. 
},label={lst:task-process-hook},firstline=130,lastline=170]
/* Implements TaskProcess() */
int TaskProcess(pool *p, task *t) {
  double state[24];
  double result[1][1][MAX_SNAPSHOTS][2];
  double tstart = 0.0, tend = 0.0, step = 0.0;
  double megno = 0.0, err = 0.0, eps = 0.0;
  int snapshots, status = SUCCESS;

  MReadOption(p, "task-checkpoints", &snapshots);
  MReadOption(p, "eps", &eps);
  MReadOption(p, "step", &step);

  // power-based time intervals
  if (t->cid == 0) tstart = 0.0;
  else tstart = pow(10, t->cid + 4);
  
  tend = pow(10, t->cid + 5);

  step = step*(pow(5,0.5)-1)/2.0;

  MReadData(t, "state", &state[0]);

  if (t->cid > 0) {
    MReadData(t, "result", &result[0][0][0][0]);
  }

  // Numerical integration goes here
  status = saba3(state, step, tstart, tend, 
      eps, &megno, &err);
  if (status < 0) return MODULE_ERR_OTHER;
  
  // Assign the master result
  result[0][0][t->cid][0] = megno;
  result[0][0][t->cid][1] = err;

  MWriteData(t, "result", &result[0][0][0][0]);
  MWriteData(t, "state", &state[0]);

  if (t->cid + 1 == snapshots) return TASK_FINALIZE;
  return TASK_CHECKPOINT;
}
\end{lstlisting}

\begin{lstlisting}[caption={\mcap{The module compilation.} The code of the module must be
compiled to a shared library with \code{libmechanic\_module\_} prefix. The module should
be linked with the \code{mechanic} library and available in the user's
\code{LD\_LIBRARY\_PATH}.
},label={lst:module-compilation},firstline=2,lastline=4]
	mpicc -fPIC -Dpic -shared aweb.c \
  	-o libmechanic_module_aweb.so \
  	-lmechanic -lm -lhdf5 
\end{lstlisting}

\begin{lstlisting}[caption={\mcap{Using the module.} The \mechanic{} should be invoked
with \code{mpiexec} or \code{mpirun} scripts. The user's module is specified  with the
\code{-p} (\code{--module}) option. In the example below, the \code{aweb} module is
invoked on the 2048 CPUs in the default, master--worker mode. The total resolution of the
simulation is $2048\times2048$ points. The result of the simulation will be stored in the
master file \code{arnold-web-master-00.h5}. The framework ships with the predefined
options helpful for conducting computations, such as run-time resolution or number of
checkpoints. The full list of options is given with the \code{--help} switch.
},label={lst:module-usage},firstline=6,lastline=12]
  mpirun -np 2048 mechanic -p aweb \
    --xres=2048 --yres=2048 \
    --xmin=-0.5 --xmax=1.5  \
    --ymin=-0.5 --ymax=1.5  \
    --step=0.5 --eps=0.04   \
    --task-checkpoints=5    \
    --name=arnold-web
\end{lstlisting}

\end{document}